\documentclass[nofootinbib,prl,superscriptaddress,a4paper,twocolumn]{revtex4-2}
\usepackage{graphicx}% Include figure files
\usepackage{dcolumn}% Align table columns on decimal point
\usepackage{bm,bbm}% bold math
\usepackage{xcolor}%colours
\usepackage{tcolorbox}
\usepackage{algorithm}
\usepackage{algpseudocode}
\usepackage{amsmath}
\usepackage{subfigure}
\usepackage{booktabs}
\usepackage{threeparttable}

\allowdisplaybreaks[4]
\usepackage[colorlinks=true,linkcolor=blue,citecolor=magenta,urlcolor=blue]{hyperref}
\makeatletter
\newcommand\org@hypertarget{}
\let\org@hypertarget\hypertarget
\renewcommand\hypertarget[2]{%
  \Hy@raisedlink{\org@hypertarget{#1}{}}#2%
  }
\makeatother
\newcommand{\ket}[1]{\left\vert#1\right\rangle}

\newcommand{\ketbra}[2]{| #1\rangle \langle #2|}

\begin{document}
%\title{Network nonlocality with quantum Elegant Joint Measurements}
%\title{Elegant Joint Measurements for quantum network correlations}
%\title{Elegant Joint Measurements for entanglement swapping and quantum correlations}
\title{Entanglement swapping and quantum correlations via Elegant Joint Measurements}

\author{Cen-Xiao Huang}
\email{These two authors contributed equally to this work.}
\affiliation{CAS Key Laboratory of Quantum Information, University of Science and Technology of China, Hefei, 230026, China}
\affiliation{CAS Center For Excellence in Quantum Information and Quantum Physics, University of Science and Technology of China, Hefei, 230026, China}
\affiliation{Hefei National Laboratory, Hefei, 230088, China}

\author{Xiao-Min Hu}
\email{These two authors contributed equally to this work.}
\affiliation{CAS Key Laboratory of Quantum Information, University of Science and Technology of China, Hefei, 230026, China}
\affiliation{CAS Center For Excellence in Quantum Information and Quantum Physics, University of Science and Technology of China, Hefei, 230026, China}
\affiliation{Hefei National Laboratory, Hefei, 230088, China}

\author{Yu Guo}
\affiliation{CAS Key Laboratory of Quantum Information, University of Science and Technology of China, Hefei, 230026, China}
\affiliation{CAS Center For Excellence in Quantum Information and Quantum Physics, University of Science and Technology of China, Hefei, 230026, China}
\affiliation{Hefei National Laboratory, Hefei, 230088, China}

\author{Chao Zhang}
\affiliation{CAS Key Laboratory of Quantum Information, University of Science and Technology of China, Hefei, 230026, China}
\affiliation{CAS Center For Excellence in Quantum Information and Quantum Physics, University of Science and Technology of China, Hefei, 230026, China}
\affiliation{Hefei National Laboratory, Hefei, 230088, China}

\author{Bi-Heng Liu}
\email{bhliu@ustc.edu.cn}
\affiliation{CAS Key Laboratory of Quantum Information, University of Science and Technology of China, Hefei, 230026, China}
\affiliation{CAS Center For Excellence in Quantum Information and Quantum Physics, University of Science and Technology of China, Hefei, 230026, China}
\affiliation{Hefei National Laboratory, Hefei, 230088, China}

\author{Yun-Feng Huang}
\affiliation{CAS Key Laboratory of Quantum Information, University of Science and Technology of China, Hefei, 230026, China}
\affiliation{CAS Center For Excellence in Quantum Information and Quantum Physics, University of Science and Technology of China, Hefei, 230026, China}
\affiliation{Hefei National Laboratory, Hefei, 230088, China}

\author{Chuan-Feng Li}
\email{cfli@ustc.edu.cn}
\affiliation{CAS Key Laboratory of Quantum Information, University of Science and Technology of China, Hefei, 230026, China}
\affiliation{CAS Center For Excellence in Quantum Information and Quantum Physics, University of Science and Technology of China, Hefei, 230026, China}
\affiliation{Hefei National Laboratory, Hefei, 230088, China}

\author{Guang-Can Guo}
\affiliation{CAS Key Laboratory of Quantum Information, University of Science and Technology of China, Hefei, 230026, China}
\affiliation{CAS Center For Excellence in Quantum Information and Quantum Physics, University of Science and Technology of China, Hefei, 230026, China}
\affiliation{Hefei National Laboratory, Hefei, 230088, China}

\author{Nicolas Gisin}
\affiliation{Group of Applied Physics, University of Geneva, 1211 Geneva 4, Switzerland}
\affiliation{Schaffhausen Institute of Technology - SIT, Geneva, Switzerland}

\author{Cyril Branciard}
\affiliation{Univ. Grenoble Alpes, CNRS, Grenoble INP, Institut N\'eel, 38000 Grenoble, France}

\author{Armin Tavakoli}
\email{armin.tavakoli@oeaw.ac.at}
\affiliation{Institute for Quantum Optics and Quantum Information -- IQOQI Vienna, Austrian Academy of Sciences, Boltzmanngasse 3, 1090 Vienna, Austria}
\affiliation{Atominstitut,  Technische  Universit{\"a}t  Wien, Stadionallee 2, 1020  Vienna,  Austria}

\date{\today}

\begin{abstract}
We use hyper-entanglement to experimentally realize deterministic entanglement swapping based on quantum Elegant Joint Measurements. These are joint projections of two qubits onto highly symmetric, iso-entangled, bases. We report measurement fidelities no smaller than $97.4\%$.  We showcase the applications of these measurements by using the entanglement swapping procedure to demonstrate quantum correlations in the form of proof-of-principle violations of both bilocal Bell inequalities and more stringent correlation criteria corresponding to full network nonlocality. Our results are a foray into entangled measurements and nonlocality beyond the paradigmatic Bell state measurement and they show the relevance of more general measurements in entanglement swapping scenarios. 
\end{abstract}

\maketitle

\textit{Introduction.---} Entangled measurements, i.e.~projections of several qubits onto a basis of entangled states, are an indispensable resource for quantum information processing. They are crucial for paradigmatic protocols such as teleportation \cite{Teleportation1993}, dense coding \cite{DenseCoding1992}, entanglement swapping \cite{Zukowski1993} and quantum repeaters \cite{Briegel1998, Duan2001}, as well as for emerging topics such as network nonlocality \cite{Review2021} and entanglement-assisted quantum communications \cite{Pauwels2021a, Pauwels2021b}.

While the entanglement of quantum states today is well-researched and is  known to have a broad fauna~\cite{Horodecki2009}, advances in the complementary case of entangled measurements has been largely focused on the paradigmatic Bell state measurement, i.e~the projection of (say) two qubits onto the four maximally entangled states $\ket{\phi^\pm}=\frac{1}{\sqrt{2}}\left(\ket{00}\pm\ket{11}\right)$ and $\ket{\psi^\pm}=\frac{1}{\sqrt{2}}\left(\ket{01}\pm\ket{10}\right)$. This measurement has been experimentally realized in a variety of contexts within the broader area of entanglement swapping and quantum correlations (see e.g.~\cite{Boschi1998, Pan1998,Jennewein2001, Yang2006, Riedmatten2005, Halder2007, Kaltenbaek2009, Schmid2009, Takeda2015, Williams2017, Guccione2020}). However, not much is known about the foundational relevance, practical implementation and overall usefulness of more general entangled measurements. 

Recently, a class of entangled two-qubit measurements has been proposed that is qualitatively different from the Bell state measurement. It displays elegant and natural symmetries and it is gaining an increasingly relevant role as a quantum information resource. These, so-called Elegant Joint Measurements (EJMs), are composed of a basis of iso-entangled states with the property that if either qubit is lost, the four possible remaining single-qubit states form a regular tetrahedron inside the Bloch sphere. Although originally introduced in the context of collective spin measurements \cite{Massar1995,  Gisin1999}, they were re-introduced in order to remedy the shortcomings of the Bell state measurement in triangle-nonlocality \cite{Gisin2019} and they were subsequently found to be connected to quantum state discrimination \cite{Czartowski2021}. Very recently, they have been used as the central component of both network nonlocality protocols, that bear no resemblance to standard Bell inequality violations \cite{Tavakoli2021}, and full network nonlocality protocols, which constitute a stronger, more genuine, notion of nonlocality in networks \cite{Pozas2022}. The progress has also motivated recent experiments that realize one type of EJM on a superconducting quantum processor \cite{Baumer2021} and as a photonic quantum walk \cite{Feng2020}.

Here, we go beyond the Bell state measurement and experimentally demonstrate entanglement swapping and quantum correlations based on the EJMs. We use hyper-entanglement between the polarization and path degrees of freedom in a pair of photons to create two pairs of maximally entangled states. Then, we realize a generic quantum circuit for implementing any EJM and demonstrate entanglement swapping, reporting high fidelities for both the measurement and the state swapped into the two initially independent qubits. We leverage these tools for tests of quantum  correlations in entanglement swapping scenarios, originally developed in \cite{Tavakoli2021, Pozas2022}, that for the first time are not based on the Bell state measurement. These tests, that may be viewed as proof-of-principle tests of quantum networks, are centered about the independence of the two entangled pairs.% in the entanglement swapping scenario.  Specifically, we both violate the bilocal Bell inequality introduced in Ref.~\cite{Tavakoli2021} and the inequalities derived in Ref.~\cite{Pozas2022} to test full network nonlocality.

\textit{Theoretical background.---} An EJM, labeled by a parameter $\theta\in[0,\frac{\pi}{2}]$, is a projection onto the following basis of a two-qubit Hilbert space \cite{Tavakoli2021}:
\begin{align}\nonumber
& \ket{\psi_{+++}^\theta}=\frac{1}{2}\left(e^{-\frac{i\pi}{4}}\ket{00}-r_+^\theta\ket{01}-r_-^\theta\ket{10}+e^{-\frac{3i\pi}{4}}\ket{11}\right),\\\nonumber
& \ket{\psi_{+--}^\theta}=\frac{1}{2}\left(e^{\frac{i\pi}{4}}\ket{00}+r_-^\theta\ket{01}+r_+^\theta\ket{10}+e^{\frac{3i\pi}{4}}\ket{11}\right),\\\nonumber
& \ket{\psi_{-+-}^\theta}=\frac{1}{2}\left(e^{-\frac{3i\pi}{4}}\ket{00}+r_-^\theta\ket{01}+r_+^\theta\ket{10}+e^{-\frac{i\pi}{4}}\ket{11}\right),\\\label{EJM}
& \ket{\psi_{--+}^\theta}=\frac{1}{2}\left(e^{\frac{3i\pi}{4}}\ket{00}-r_+^\theta\ket{01}-r_-^\theta\ket{10}+e^{\frac{i\pi}{4}}\ket{11}\right),
\end{align}
where $r_\pm^\theta=\frac{1\pm e^{i\theta}}{\sqrt{2}}$.
We denote the four possible outcomes of the measurement (indicated as the states' subscripts) by a string of three bits $b=(b^1,b^2,b^3)\in\{+1,-1\}^3$ such that $b^1b^2b^3=1$. The elegant property of these measurements is that all basis states are equally entangled and that the two sets of four reduced states, when the right or left qubit is lost respectively, form two mirror-image regular tetrahedra (of radius $\frac{\sqrt{3}}{2}\cos\theta$) inside the Bloch sphere whose vertices are parallel and anti-parallel with the Bloch sphere direction $(b^1,b^2,b^3)$ respectively. %The degree of entanglement in the basis states \eqref{EJM}  increases  with $\theta$, and thus the radii of the local tetrahedra decrease with $\theta$.
Notably, for $\theta=\frac{\pi}{2}$, the EJM is equivalent to the Bell state measurement up to local unitaries. 

We apply the EJM for entanglement swapping. Consider that qubits $B_1$ and $B_2$ in the two, initially independent, maximally entangled states $\ket{\phi^+}_{AB_1}\otimes \ket{\phi^+}_{B_2C}$ are subjected to the EJM. This produces the output $b$ with probability $p(b)=\frac{1}{4}$ and stochastically renders qubits $A$ and $C$ in one of the four iso-entangled states (up to complex conjugation). Consider now that the qubits $A$ and $C$ can each be independently measured  with the three Pauli observables (sometimes up to a sign), specifically $-\sigma_X$, $\sigma_Y$ and $-\sigma_Z$. For qubit $A$ ($C$) we associate these to inputs labeled $x\in\{1,2,3\}$ ($z\in\{1,2,3\}$) and label the outputs $a\in\{\pm1\}$ ($c\in\{\pm1\}$). Examining the correlators between the three measurement events, one finds that $\langle A_xB^yC_z\rangle=-\frac{1+(-1)^\sigma \sin\theta}{2}$, where $\sigma=0$ ($\sigma=1$) if $(x,y,z)$ is an even (odd) permutation of $(1,2,3)$, and $\langle A_xB^yC_z\rangle=0$ otherwise. Moreover, the two-body correlators are $\langle A_xB^y\rangle=-\frac{\cos\theta}{2}\delta_{x,y}$, $\langle B^yC_z\rangle=\frac{\cos\theta}{2}\delta_{y,z}$ (where $\delta$ is the Kronecker delta) and $\langle A_xC_z\rangle=0$, and the one-body correlators all vanish. Here, the correlators are defined as $\langle A_xB^yC_z\rangle=\sum_{a,b,c} ab^yc \hspace{1mm}p(a,b,c|x,z)$ and analogously for the two- and one-body cases.

In Ref.~\cite{Tavakoli2021} it was shown that the above quantum correlations cannot be modelled by any bilocal hidden variable theory, i.e.~any model that ascribes an independent local variable to each of the two states associated to systems $AB_1$ and $B_2C$ respectively (see e.g.~\cite{Branciard2010, Branciard2012}). This is witnessed through the violation of the  bilocal Bell inequality
\begin{equation}\label{biloc}
\mathcal{B}\equiv \frac{S}{3}-T\leq 3+f(Z),    
\end{equation}
where $S=\sum_{k=1}^3\left(\langle B^kC_k\rangle-\langle A_kB^k\rangle\right)$, $T=\sum_{x\neq y\neq z\neq x}\langle A_xB^yC_z\rangle$ and $Z=\max(\mathcal{C})$ where $\mathcal{C}=\{|\langle A_1\rangle|,\ldots,|\langle A_3B^3C_3\rangle|\}$ is the list of the absolute value of all one-, two- and three-body correlators that do not appear in the definitions of $S$ or $T$. The term $f(Z)$ is a correction term relevant to the experimental reality that  measured correlators in $\mathcal{C}$ will not equal zero. In Appendix we numerically show that $f(Z) = Z + 4 Z^2$ is a valid correction term as long as $Z \lesssim 0.55$. The  quantum protocol achieves $\mathcal{B}=3+\cos\theta$, which for an ideal implementation ($Z=0$) gives a violation for every EJM except the Bell state measurement ($\theta=\frac{\pi}{2}$). In contrast to many other criteria for network nonlocality, which are tailored for employing the Bell state measurement (see e.g.~\cite{Branciard2010, Branciard2012, Tavakoli2017, Tavakoli2014, Gisin2017, Andreoli2017}), the quantum protocol and bilocal Bell inequality are not based on using standard Bell nonlocality as a building block for network nonlocality.

\begin{figure}
\includegraphics[width=0.45\textwidth]{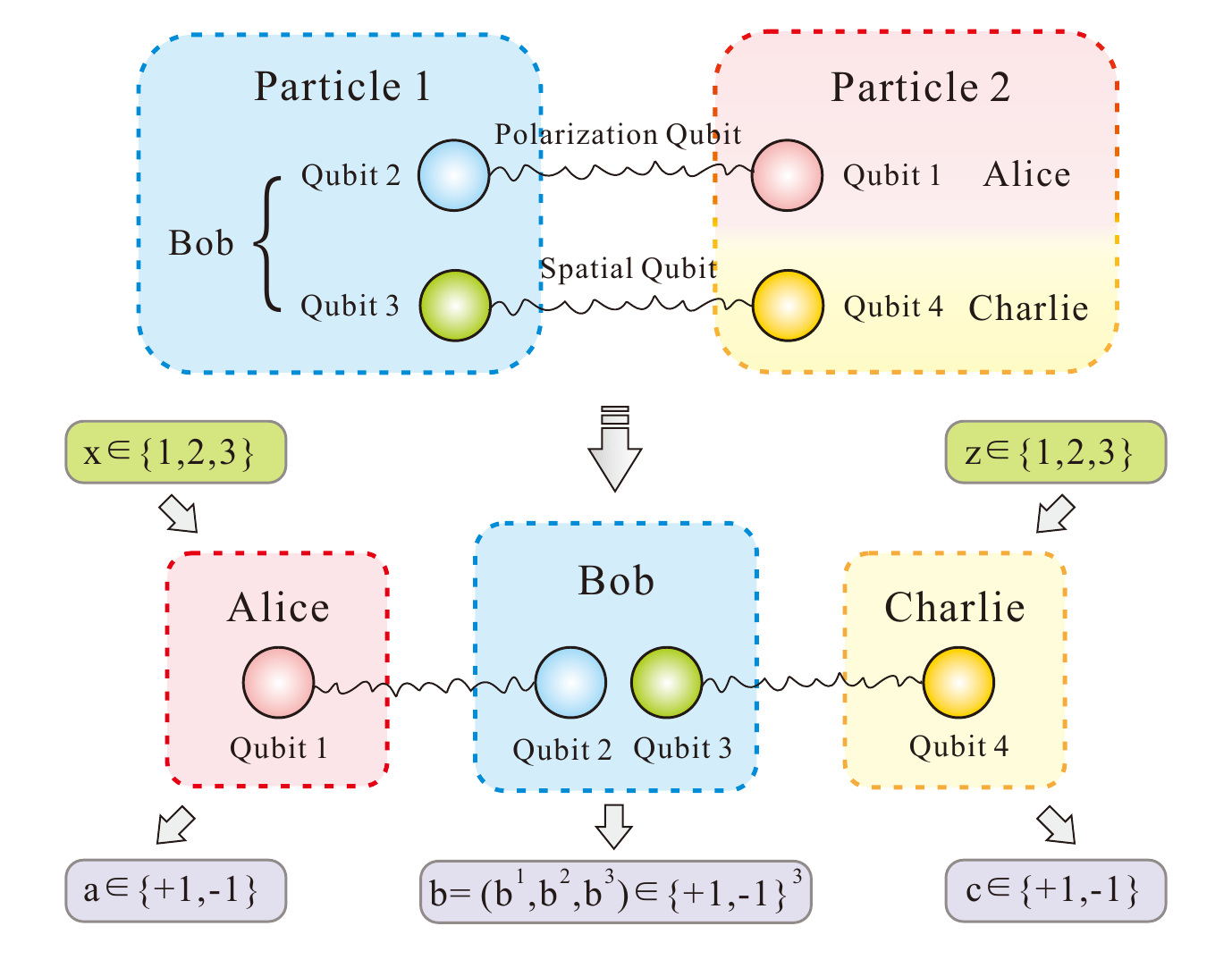}
\vspace{-0.5cm}
\caption{Schematic diagram. Particles 1 and 2 are in a hyper-entangled state $|\phi\rangle=|\phi^{+}_p\rangle_{12}\otimes|\phi^{+}_s\rangle_{34}$ of polarisation qubits (p) and spatial qubits (s). We assign the state to Alice, Bob and Charlie and let Bob perform the EJM on qubits 2 and 3, and let Alice and Charlie measure  $\{\sigma_X, \sigma_Y, \sigma_Z\}$ on the  qubits 1 and 4 respectively.
%where qubits 1 and 2 are polarization qubits, while qubits 3 and 4 are spatial qubits. In order to test for non-bilocality and full network nonlocality, we assign this state to three parties: Alice, Bob and Charlie---and thus write it as $|\phi\rangle=|\phi^{+}\rangle_{AB_1}\otimes|\phi^{+}\rangle_{B_2C}$. Via the inputs  $x,z\in\{1,2,3\}$, Alice and Charlie choose different measurement settings $\{\sigma_X, \sigma_Y, \sigma_Z\}$ with binary outcomes $a,c\in\{+1, -1\}$. Bob performs an EJM and has a four-valued outcome $b=(b^1,b^2,b^3)\in\{+1, -1\}^3$. \armin{Do we really need to keep this figure? Its content is pretty much explained in the main text and we are still about 150 words above the prl limit. Alternatively, we could cut parts of the caption because it is repeating what it is in the main text.}
}
\label{fig1}
\end{figure}

\begin{figure*}[t!]
\includegraphics[width=0.9\textwidth]{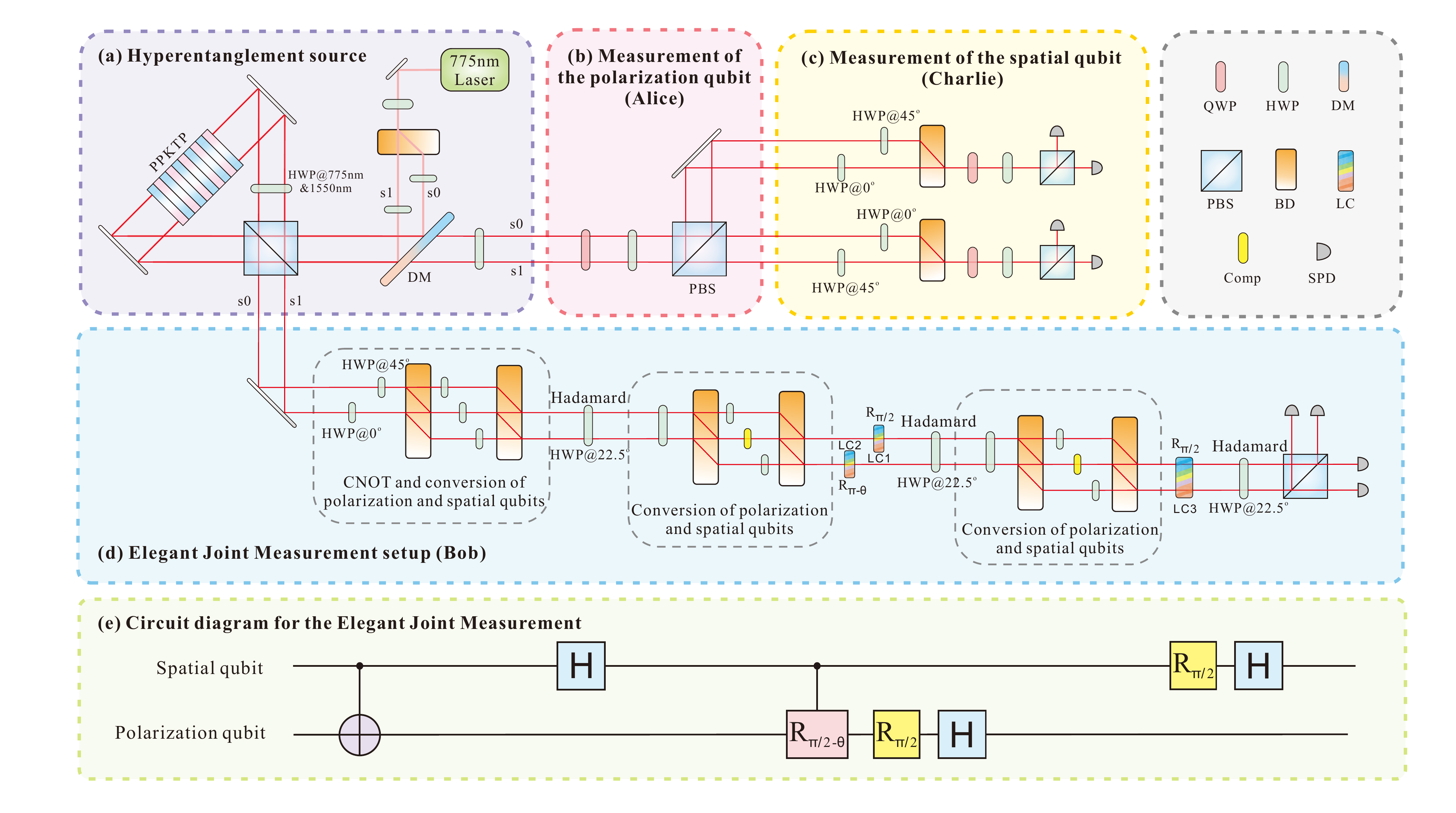}
\vspace{-0.5cm}
\caption{Experimental setup. (a) Preparation of the hyper-entangled states $\frac{1}{2}(|H\rangle|H\rangle+|V\rangle|V\rangle)\otimes(|s_0\rangle|s_0\rangle+|s_1\rangle|s_1\rangle)$.
(b) Alice's measurement of the polarization qubit. The quarter and half wave plates (QWP, HWP) are used to set the three possible projection measurements $\{\sigma_X,\sigma_Y,\sigma_Z\}$. (c) Charlie's measurement of the spatial qubit (first converting it into polarisation). (d) Bob's EJM setup. We use different optical elements on the two paths (combined appropriately with the subsequent polarization-to-spatial conversion setup, see details in Appendix) to realize the CNOT gate.
The Hadamard operation on the spatial qubit is realized by a HWP set at $22.5^{\circ}$, between two polarization-to-spatial conversion setups. Then two liquid crystals (LC) in the different paths and a HWP at $22.5^{\circ}$ are used to realize the C-phase, $\frac{\pi}{2}$ phase and Hadamard gates on the polarization qubit, where the voltage of LC1 (LC2) is set so that it realizes $R_{\frac{\pi}{2}}$ ($R_{\pi-\theta}= R_{\frac{\pi}{2}-\theta}\cdot R_{\frac{\pi}{2}}$, resp.) with $R_\phi = \left(\begin{smallmatrix}
    1 & 0 \\
    0 & e^{i\phi}
\end{smallmatrix}\right)$. Finally, we convert the spatial qubit to a polarization qubit again, and use LC3 and HWP to realize the $\frac{\pi}{2}$ phase and Hadamard gates on the spatial (turned into polarization) qubit. (e) Circuit diagram for the EJM \cite{Tavakoli2021}. DM, dichroic mirror; PBS, polarization beam splitter; Comp, compensator; SPD, single photon detector.}
\label{fig2}
\end{figure*}

The quantum correlations also reveal stronger forms of network nonlocality. Ref.~\cite{Pozas2022} introduced the concept of full network nonlocality, which constitutes a more genuine  network phenomenon.  Again assuming only the initial independence of systems $AB_1$ and $B_2C$, the correlations are said to be full network nonlocal if they cannot be modelled by any theory in which one source corresponds to a local variable and the other to a generalized, perhaps even post-quantum, nonlocal resource. Notably, many known network Bell inequalities, tailored for the Bell state measurement, fail to reveal full network nonlocality \cite{Review2021}. % In particular, this includes the most well-known and experimentally most well-tested bilocal Bell inequality . 

However, the EJMs enable a successful detection. Full network nonlocality is implied by the simultaneous violation of both the following inequalities \cite{Pozas2022}:
\begin{equation}\label{fnn1}
\begin{aligned}
\mathcal{F}_{1}=&-\left\langle A_{1} B^{2} C_{3}\right\rangle-\left\langle A_{2} B^{2}\right\rangle \cr
&+\left\langle C_{3}\right\rangle\left[\left\langle A_{1} B^{2}\right\rangle+\left\langle A_{2} B^{2} C_{3}\right\rangle+\left\langle C_{3}\right\rangle\right] \leq 1 ,
\end{aligned}
\end{equation}
\begin{equation}\label{fnn2}
\begin{aligned}
\mathcal{F}_{2}=&-\left\langle A_{1} B^{2} C_{3}\right\rangle+\left\langle B^{2} C_{2}\right\rangle \cr
&+\left\langle A_{1}\right\rangle\left[\left\langle B^{2} C_{3}\right\rangle-\left\langle A_{1} B^{2} C_{2}\right\rangle+\left\langle A_{1}\right\rangle\right] \leq 1 .
\end{aligned}
\end{equation}
The given quantum protocol achieves $\mathcal{F}_1=\mathcal{F}_2=\frac{1}{2}\left(1+\sin\theta+\cos\theta\right)$, which is a violation for every $\theta\in (0,\frac{\pi}{2})$. The largest violations are obtained for an intermediate member of the EJM family, namely $\theta=\frac{\pi}{4}$. Notice that these violations are achieved using effectively only a binarised version of the EJM, as only Bob's outcome $B^2$ appears in the inequalities above. Interestingly, and in contrast to the Bell state measurement, it remains entangled even after binarisation. %which has no analogy with the Bell state measurement as binarisation makes the latter separable and hence resourceless for entanglement swapping.

\begin{figure*}[t]
\centering
\subfigure{
\includegraphics[width=0.46\textwidth]{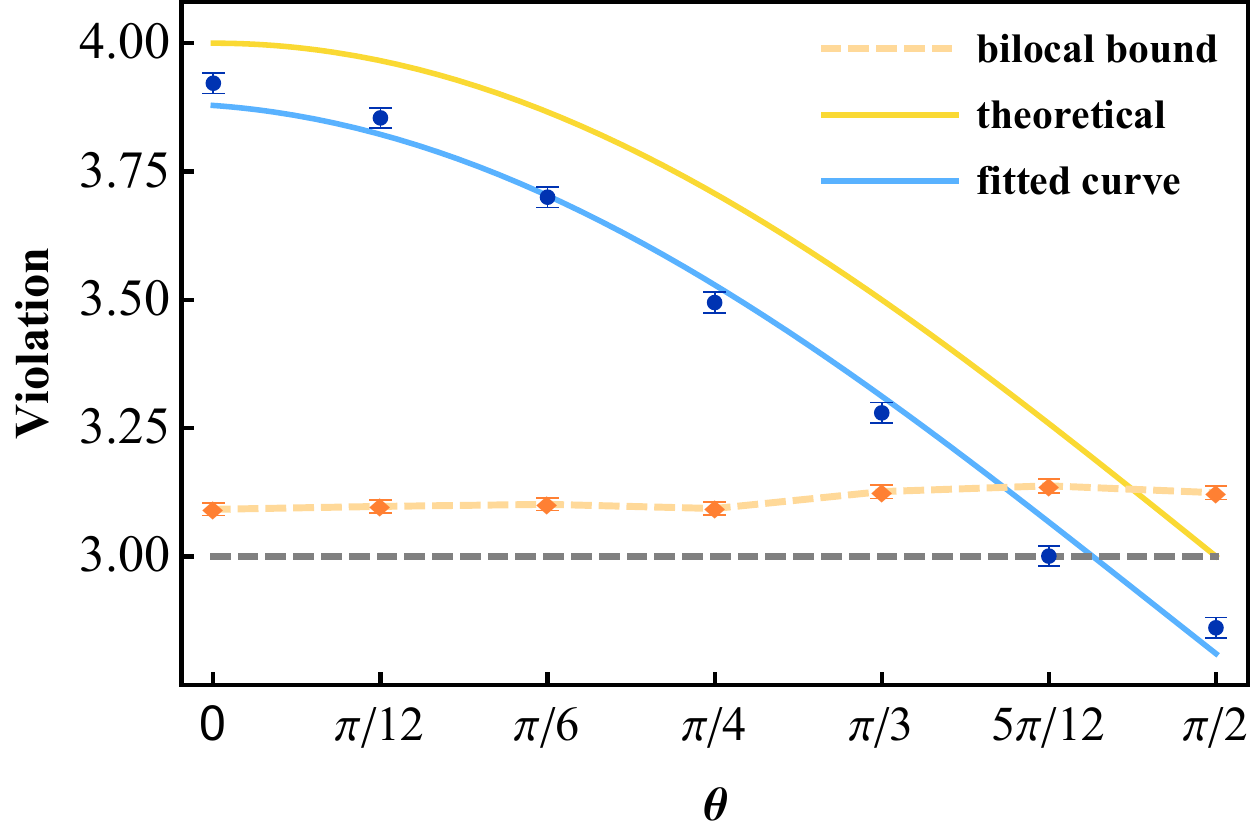}
}
\subfigure{
\includegraphics[width=0.46\textwidth]{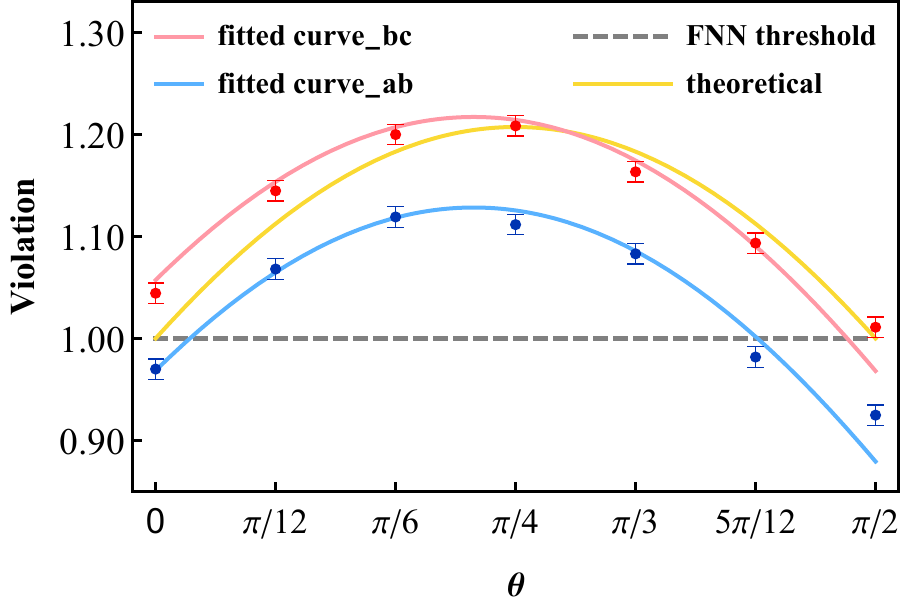}
}
\caption{Experimental results. (Left) Results for our test of the bilocal Bell inequality~\eqref{biloc}. The yellow curve is the theoretical prediction for the quantum correlations under consideration. The blue points are the experimental data and the blue line is a fitted curve. The orange diamonds are the bilocal bounds after consideration of the $Z$-correction term. (Right) Results for our test of full network nonlocality, via inequalities~\eqref{fnn1}--\eqref{fnn2}. The yellow curve is the theoretical prediction. The blue and red points are the experimentally measured values corresponding to $\mathcal{F}_1$ and $\mathcal{F}_2$ respectively. The errors were estimated assuming Poissonian statistics.}
\label{fig3}
\end{figure*}

\textit{Experimental setup.---}
Our approach to experimentally realize the EJM and the associated tests of quantum correlations is represented in Fig.~\ref{fig1}. A central fact is that in linear optics, EJMs cannot be realized without auxiliary particles when each qubit is encoded onto a different photon~\cite{Yoran, Lutkenhaus1999,Loock2004,Tavakoli2021}. Our approach is therefore to circumvent this issue by using two different degrees of freedom, path and polarization, of the same photon. 

We generate pairs of hyper-entangled states $|\phi\rangle=|\phi^{+}_p\rangle_{12}\otimes|\phi^{+}_s\rangle_{34}$. 
Here, $|\phi^{+}_p\rangle_{12}=\frac{1}{\sqrt{2}}(|H\rangle|H\rangle+|V\rangle|V\rangle)$ is a polarization Bell state of qubits 1 and 2, and $|\phi^{+}_s\rangle_{34} =\frac{1}{\sqrt{2}}(|s_0\rangle|s_0\rangle+ |s_1\rangle|s_1\rangle)$ is a spatial mode Bell state of qubits 3 and 4. Fig.~\ref{fig2}a illustrates the polarization-entangled photon pair production via a type-II cut periodically poled potassium titanyl phosphate crystal. Using the first beam displacer (BD), we split the pump laser (775nm) to two spatial modes ($s_{0}$ and $s_{1}$) and generate polarization-entangled photon pair in each mode. Thus, we generate the hyper-entanglement $|\phi\rangle=|\phi^{+}_p\rangle_{12}\otimes|\phi^{+}_s\rangle_{34}$ by tuning the relative phase between the two spatial modes \cite{Hu2021,HUANG2021}. We used
90 mW pumped light to excite about 2000 photon pairs per second. The ratio of coincidence counts to single counts in the entangled source is $19\%$. Qubits 1 and 2  are encoded in the polarization and path degrees of freedom of particle 1, and qubits  3 and 4 are encoded in the polarization and path degrees of freedom of particle 2. Attributing qubit 1 to Alice, qubits 2 and 3 to Bob, and qubit 4 to Charlie, we can rewrite the prepared state as: $|\phi\rangle=|\phi^{+}\rangle_{AB_1}\otimes|\phi^{+}\rangle_{B_2C}$.

Fig.~\ref{fig2}d illustrates our deterministic implementation of EJMs on the qubits 2 (polarization) and 3 (spatial mode). This is based on the quantum circuit proposed in Ref.~\cite{Tavakoli2021}: it requires CNOT, C-Phase, Phase and Hadamard operations (see Fig.~\ref{fig2}e).
%We choose the spatial degree of freedom to control the polarization, and the CNOT gate is realized by a half wave plate (HWP) in different paths. 
We choose the spatial degree of freedom to control the polarization, so that the controlled gates can be realized using different optical elements (acting on the polarisation) on the different paths (see Appendix for details). 
The CNOT gate is combined with the conversion part and is realized by setting HWPs at different degrees on the paths $s_0$ and $s_1$.
%When the \huang{spatial mode is $s_1$ ($s_0$)}, the polarization is (not) flipped. 
A similar C-phase gate is realized by a liquid crystal phase plate. Loading different voltages on the liquid crystal (LC)  produces different phases between horizontally ($H$) polarized light and vertically ($V$) polarized light. In the case of path $s_0$, we do not change the phase between $H$ and $V$. In the case of path $s_1$, we change the phase between $H$ and $V$ to realize the C-Phase gate.
%By changing the voltage of LC, we can load any possible phase in the EJM with high fidelity. 
We use the liquid phase crystal to load $\pi/2$ phase on $H$ and $V$ to complete the $\pi/2$ phase gate of the polarization qubit. The Hadamard gate acting on the polarization qubit is realized by setting a half wave plate at 22.5 degrees. We realize the phase gate and Hadamard gate by converting the path qubit into a polarization qubit. By cascading these gate operations, we realize the EJM  on polarization and path qubits of a single photon.

Finally, we check for correlations between the initially independent polarization and path qubits, 1 and 4, by measuring $\{\sigma_X, \sigma_Y, \sigma_Z\}$ on both sides, as illustrated in Fig.~\ref{fig2}b and c.

\textit{Experimental results.---} We have implemented eight different choices of the EJM parameter $\theta\in \{0,\frac{\pi}{12},\frac{\pi}{6},\frac{\pi}{4},\frac{\pi}{3},\frac{5\pi}{12},\frac{\pi}{2}\}$. We reconstruct our quantum measurement from the obtained data via measurement tomography following the maximum-likelihood method~\cite{PhysRevA.64.024102}. In particular, we record a measurement fidelity of  $98.5\pm0.1\%$ for $\theta=0$ and $97.5\pm0.2\%$ for $\theta=\frac{\pi}{4}$. Details are provided in  Appendix.  Moreover, we have measured the fidelity of our entanglement swapping procedure through the fidelity between the EJM eigenstates and the post-measurement state of system $AC$. For the two most relevant cases, namely $\theta=0$ and $\theta=\frac{\pi}{4}$, the average fidelity of the swapped state is $98.5\pm0.2\%$ and $97.5\pm0.2\%$.

For each of the chosen values of $\theta$, we have tested the bilocal Bell inequality \eqref{biloc} and the criterion (\ref{fnn1}-\ref{fnn2}) for full network nonlocality. For each setting $(x,z)$ we measure for ten seconds, recording approximately 20000 coincidences. We observe correlations strong enough to demonstrate both non-bilocality and full network nonlocality. For the former, we obtain the largest violation by implementing the EJM at  $\theta=0$, measuring $\mathcal{B}=3.922\pm0.018$ while the right hand side of inequality~\eqref{biloc} (its bilocal bound) is $3.092\pm 0.012$. For the latter, we obtain at best $\mathcal{F}_{1}=1.112\pm0.006$ and $\mathcal{F}_{2}=1.208\pm0.006$ by choosing $\theta=\frac{\pi}{4}$. We note that our data, for $\theta=0$, also provides  a violation of the second bilocal Bell inequality originally proposed in Ref.~\cite{Tavakoli2021}, specifically achieving  $\mathcal{B^{'}}=29.333\pm0.019$, which exceeds the bilocal bound of $28.531$ (see Appendix for details). 

%corresponding to a violation of inequality (equation (3)) of almost 20 sigmas.

Our complete correlation results are illustrated in Fig.~\ref{fig3}. For the bilocality test, we measured $Z=0.071 \pm 0.008$ for the most interesting case of $\theta=0$ and at most $Z=0.099 \pm 0.008$ over all $\theta$. Taking into account the $Z$-dependent correction to the bilocal bound in \eqref{biloc}, we record a violation for the first five values of $\theta$. In addition, we observe full network nonlocality for four different choices of $\theta$. Although the theory predicts $\mathcal{F}_1=\mathcal{F}_2$, we consistently find that $\mathcal{F}_2$ is significantly larger than $\mathcal{F}_1$. This is attributed to the phase error in the EJM setup. As discussed in Appendix, a small amount of such error induces a considerable offset in the values of $\mathcal{F}_1$ and $\mathcal{F}_2$. 

These correlation tests are based on the assumption of independent entangled pairs. To reasonably meet this assumption, we have carefully calibrated our setup in order to eliminate potential correlations between the initial system $AB_1$ and $B_2C$. In order to estimate the accuracy of the state preparation we have, via state tomography, reconstructed the total state and found that it has a fidelity of $99.0\pm0.1\%$ with the target state $|\phi^{+}\rangle_{AB_1}\otimes|\phi^{+}\rangle_{B_2C}$. To estimate the correlations between the two joint systems, we have evaluated 
both the fidelity and the quantum mutual information between the tomographic reconstruction and the product of its reductions to systems $AB_1$ and $B_2C$. We obtain $99.1\pm 0.1\%$ and $0.048\pm 0.003$ bits respectively.

\textit{Discussion.---} %We have reported high-fidelity  deterministic implementations of entanglement swapping and quantum correlations based on EJMs. 
Our work constitutes a first step towards the experimental realisation of entanglement swapping protocols beyond the celebrated Bell state measurement, and our experiments showcase their advantages. On the conceptual side, it is interesting to further understand the role of more general entangled measurements in quantum information processing. Already conceptualising the extension of the EJM to multipartite settings appears to not be straightforward. On the technological side, a natural next step is to investigate entanglement swapping tests based on EJMs where all qubits are assigned separate optical carriers, i.e.~with the help of auxiliary particles or nonlinear optical processes. This would enable the use of deterministic and complete EJMs in proper quantum networks. Also, provided appropriate theoretical advances take place (see e.g.~\cite{Krivachy2020}), it may be interesting to extend our  hyperentanglement-based approach towards proof-of-principle  demonstrations of triangle-nonlocal correlations via EJMs \cite{Gisin2019}.

\begin{acknowledgments}
This work was supported by the National Key Research and Development Program of China (2021YFE0113100 and 2017YFA0304100), the National Natural Science  Foundation of China (11874345, 11821404, 11904357, 12174367, and 12175106), the Fundamental Research Funds for the Central Universities, USTC Tang Scholarship, Science and Technological Fund of Anhui Province for Outstanding Youth (2008085J02), China Postdoctoral Science Foundation (2021M700138), China Postdoctoral for Innovative Talents (BX2021289). AT was supported by the Wenner-Gren Foundations. NG was supported by the Swiss National Science Foundation via the National Centres of Competence in Research (NCCR)-SwissMap. 
\end{acknowledgments}

%\section{REFERENCES AND NOTES}
\bibliography{bibliography}

\newpage
\onecolumngrid
\section{Appendix}
\subsection{(A) Noise analysis}
The experimental curves shown in Fig.~\ref{fig3} are obtained by numerical fitting, in which we add extra noise to the state and operation gates in EJM. We use the least squares method to determine the fit, including six independent parameters which we discuss now.

The state preparation is imperfect. We assume here that the input state takes the form $\rho=\rho_1\otimes\rho_2$
where each system is subject to isotropic noise, i.e.~$\rho_i=V_i |\phi^+\rangle \langle \phi^+|+\frac{1-V_i}{4} I$. In our noise model we set $V=V_1=V_2$, as we observe experimentally that the two visibilities are indeed very close.

\begin{figure}[tph!]
\includegraphics[width=0.9\textwidth]{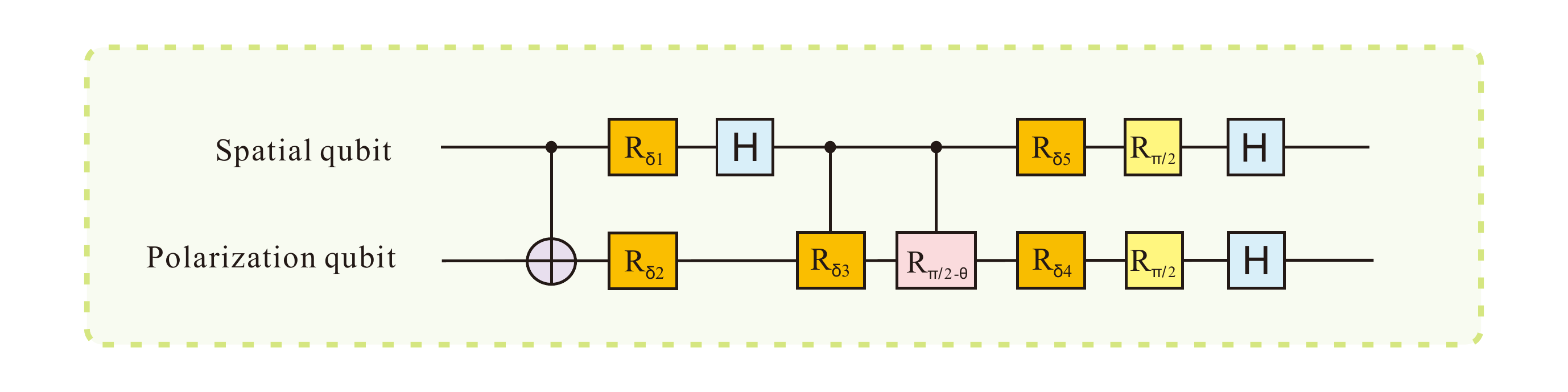}
\vspace{-0.5cm}
\caption{The error model for the EJM. Five phase error gates ($R_{\delta_1}-R_{\delta_5}$, orange) are added within the EJM circuit.}
\label{fig4}
\end{figure}

Then we consider an error model for the EJM. In our setup, the phase errors in the EJM gate operations have a considerable influence on the experiment. To model this, we introduce five phase error parameters to simulate the phase mismatch in the measurement as shown on Fig.~\ref{fig4}: these phase errors are added to the phase calibration locations (such as interferometers or liquid crystals), which usually produce extra phase in the experiment. For each phase gate (orange in the figure), we ascribe a phase error $e^{i\delta_n}$, that is, the gate applies the unitary operation
\begin{equation*}
R_{\delta_n}=
\left(
\begin{array}{cc}
1 & 0 \\
0 & e^{i\delta_n}
\end{array}
\right).
\end{equation*}

After adding these five error gates,  we use the least square method to numerically fit the three sets of data points in Fig.~\ref{fig3}. By calculating the minimum square error of the fitting function and data points, we can obtain the curves in Fig.~\ref{fig3}. The result is $\{V\rightarrow 0.984, \delta_1 \rightarrow 0, \delta_2 \rightarrow 0.022\pi, \delta_3\rightarrow -0.033\pi, \delta_4\rightarrow -0.004\pi, \delta_5\rightarrow -0.011\pi\}$.
In this noise model, the fidelity of the 4-qubit global state is 0.976, and the fidelity of the measurement---defined here as%
\footnote{The fidelity between two POVMs $\{E_b\}_b$ and $\{E_b'\}_b$ (with their $d$ elements in one-to-one correspondence) can be defined as the fidelity between the two states $\sigma = \frac{1}{d}\sum_b E_b \otimes |b\rangle\langle b|$ and $\sigma' = \frac{1}{d}\sum_b E_b' \otimes |b\rangle\langle b|$, where $\{|b\rangle\}_b$ is a $d$-dimensional orthonormal basis for some ancillary system~\cite{hou2018deterministic}. If one of the two POVMs is a projective measurement with rank-1 operators, as considered here, this fidelity simplifies to the expression given above.}
$F=\left(\frac{1}{4}\sum_b \sqrt{\langle\psi_b^\theta|E_b^\theta|\psi_b^\theta\rangle}\right)^2$, where $\{E_b^\theta\}_b$ is the POVM that approximates the ideal EJM projection onto the basis $\{|\psi_b^\theta\rangle\}_b$---is 0.997 (independently of $\theta$).

\bigskip

Naturally, the de-facto noise in the experiment is more complex. 
For the input state preparation, we performed 4-qubit tomography~\cite{Hu:20}, from which we evaluated the fidelity with respect to the ideal input state $|\phi^{+}\rangle_{AB_1}\otimes|\phi^{+}\rangle_{B_2C}$ to be $99.0\pm0.1\%$. For the EJM, we used
 measurement tomography to reconstruct the measurement operators~\cite{PhysRevA.64.024102}. By inputting a tomographically complete set of states, we can obtain our measurement operators through the outcome statistics. The input state are $\{|00\rangle,|01\rangle,|10\rangle,|11\rangle,\frac{1}{\sqrt{2}}|0\rangle\otimes(|0\rangle+|1\rangle),...,\frac{1}{2}(|0\rangle-i|1\rangle)\otimes(|0\rangle-i|1\rangle)\}$, giving a total of 36 tomographically complete states. For $\theta=0$, the tomographic reconstruction gives the measurement operators

\begin{equation*}
E_1^{\theta=0}=
\left(
\begin{array}{rrrr}
   \multicolumn{1}{c}{0.246} & -0.256 + 0.233i & -0.024 - 0.011i & -0.018 + 0.249i\\
  -0.256 - 0.234i &  \multicolumn{1}{c}{0.488} &  0.016 + 0.034i &  0.253 - 0.240i\\
  -0.024 + 0.011i &  0.016 - 0.034i &  \multicolumn{1}{c}{0.005} & -0.013 - 0.022i\\
  -0.018 - 0.249i &  0.253 + 0.240i & -0.013 + 0.022i &  \multicolumn{1}{c}{0.268}

\end{array}
\right)
\end{equation*}

\begin{equation*}
E_2^{\theta=0}=
\left(
\begin{array}{rrrr}
   \multicolumn{1}{c}{0.266} & -0.012 + 0.012i &  0.245 + 0.265i &  0.002 - 0.238i\\
  -0.012 - 0.012i &  \multicolumn{1}{c}{0.002} & -0.001 - 0.022i & -0.008 + 0.013i\\
   0.245 - 0.265i & -0.000 + 0.022i &  \multicolumn{1}{c}{0.493} & -0.239 - 0.227i\\
   0.002 + 0.238i & -0.008 - 0.013i & -0.239 + 0.227i &  \multicolumn{1}{c}{0.233}
\end{array}
\right)
\end{equation*}

\begin{equation*}
E_3^{\theta=0}=
\left(
\begin{array}{rrrr}
   \multicolumn{1}{c}{0.246} &  0.025 + 0.002i & -0.233 - 0.257i &  0.006 - 0.244i\\
   0.025 - 0.002i &  \multicolumn{1}{c}{0.003} & -0.025 - 0.024i & -0.003 - 0.024i\\
  -0.233 + 0.257i & -0.025 + 0.024i &  \multicolumn{1}{c}{0.500} &  0.251 + 0.243i\\
   0.006 + 0.244i & -0.003 + 0.024i &  0.251 - 0.243i &  \multicolumn{1}{c}{0.250} 
\end{array}
\right)
\end{equation*}

\begin{equation*}
E_4^{\theta=0}=
\left(
\begin{array}{rrrr}

   \multicolumn{1}{c}{0.243} &  0.243 - 0.248i &  0.012 + 0.003i &  0.010 + 0.233i\\
   0.243 + 0.248i  & \multicolumn{1}{c}{0.507} &  0.009 + 0.012i & -0.242 + 0.251i\\
   0.012 - 0.003i &  0.009 - 0.012i &  \multicolumn{1}{c}{0.002} &  0.002 + 0.007i\\
   0.010 - 0.233i & -0.242 - 0.251i &  0.002 - 0.007i &  \multicolumn{1}{c}{0.249}

\end{array}
\right)
\end{equation*}
so that the ``fidelity'' of each (nonnormalized) POVM element with that of the ideal EJM, defined here simply as $\langle\psi_b^\theta|E_b^\theta|\psi_b^\theta\rangle$, is $98.8\pm0.2\%$, $97.8\pm0.3\%$, $98.8\pm0.2\%$ and  $98.5\pm0.2\%$ respectively. On average, the measurement fidelity (as defined above) is $98.5\pm0.1\%$.

For $\theta=\frac{\pi}{4}$, the result are

\begin{equation*}
E_1^{\theta=\frac{\pi}{4}}=
\left(
\begin{array}{rrrr}
   \multicolumn{1}{c}{0.237} & -0.120 + 0.264i & -0.153 - 0.072i & -0.022 + 0.243i\\
  -0.120 - 0.264i &  \multicolumn{1}{c}{0.372} & -0.004 + 0.208i &  0.297 - 0.087i\\
  -0.153 + 0.072i & -0.004 - 0.208i &  \multicolumn{1}{c}{0.121} & -0.060 - 0.166i\\
  -0.022 - 0.243i &  0.297 + 0.087i & -0.060 + 0.166i &  \multicolumn{1}{c}{0.273}

\end{array}
\right)
\end{equation*}

\begin{equation*}
E_2^{\theta=\frac{\pi}{4}}=
\left(
\begin{array}{rrrr}
   \multicolumn{1}{c}{0.271} & -0.060 + 0.165i &  0.284 + 0.132i &  0.010 - 0.241i\\
  -0.059 - 0.165i &  \multicolumn{1}{c}{0.115} &  0.017 - 0.203i & -0.148 + 0.048i\\
   0.284 - 0.132i &  0.017 + 0.203i &  \multicolumn{1}{c}{0.373} & -0.098 - 0.265i\\
   0.010 + 0.241i & -0.148 - 0.048i & -0.098 + 0.265i &  \multicolumn{1}{c}{0.233}
\end{array}
\right)
\end{equation*}

\begin{equation*}
E_3^{\theta=\frac{\pi}{4}}=
\left(
\begin{array}{rrrr}
   \multicolumn{1}{c}{0.241} &  0.071 - 0.140i & -0.281 - 0.120i &  0.010 - 0.243i\\
   0.071 + 0.140i &  \multicolumn{1}{c}{0.108} & -0.012 - 0.208i &  0.142 - 0.062i\\
  -0.281 + 0.120i & -0.012 + 0.208i &  \multicolumn{1}{c}{0.401} &  0.102 + 0.285i\\
   0.010 + 0.243i &  0.142 + 0.062i &  0.102 - 0.285i &  \multicolumn{1}{c}{0.251}
\end{array}
\right)
\end{equation*}

\begin{equation*}
E_4^{\theta=\frac{\pi}{4}}=
\left(
\begin{array}{rrrr}
   \multicolumn{1}{c}{0.251} &  0.108 - 0.288i &  0.149 + 0.060i &  0.002 + 0.241i\\
   0.108 + 0.288i &  \multicolumn{1}{c}{0.405} & -0.001 + 0.203i & -0.291 + 0.101i\\
   0.149 - 0.060i & -0.001 - 0.203i  & \multicolumn{1}{c}{0.106} &  0.056 + 0.146i\\
   0.002 - 0.241i & -0.291 - 0.101i &  0.056- 0.146i &  \multicolumn{1}{c}{0.244}

\end{array}
\right)
\end{equation*}
where the ``fidelity'' of each (nonnormalized) POVM element with  that of the EJM is $97.4\pm0.3\%$, $96.4\pm0.3\%$, $97.8\pm0.3\%$ and $98.4\pm0.2\%$ respectively. The average measurement fidelity is $97.5\pm0.2
\%$.

\medskip

As we see, the noise model considered above gave slightly different values for the input state and measurement fidelities than those obtained directly from the tomographic reconstructions. This is not surprising, as the noise model was restricted to only a few parameters, and does not claim to faithfully describe all actual sources of errors.
Below we further investigate the effect of one of these sources of errors considered in our model onto the value of the FNN inequalities.

\subsection{(B) Influence of phase error on $\mathcal{F}_1$ and $\mathcal{F}_2$}
Although the theoretical prediction stipulates that $\mathcal{F}_1=\mathcal{F}_2$ for every $\theta$, we see in the experiment that $\mathcal{F}_2$ is significantly and consistantly larger than $\mathcal{F}_1$. We now show that a considerable factor in explaining this difference can be due to a small phase offset in the circuit for the EJM between the two qubits. Consider that the circuit is implemented accurately, with the exception of the gate $R_{\pi/2}$ on (say) the spatial qubit instead corresponding to a rotation $\ketbra{0}{0}+e^{i\phi}\ketbra{1}{1}$, where $\phi$ is only approximately $\pi/2$. In the noise model illustrated in Figure~\ref{fig4}, this corresponds to setting $\delta_1=\delta_2=\delta_3=\delta_4=0$ and considering a small but nonzero error $\delta_5 = \phi-\frac{\pi}{2}$.

For the bilocality parameter, for the most relevant case of $\theta=0$, one finds that the impact of the deviation $\delta\equiv \delta_5 = \phi-\frac{\pi}{2}$ is only relevant to second order. Specificially, up to second order about $\delta\approx 0$, one has
\begin{equation}
    \mathcal{B}\approx 4-\frac{4}{3}\delta^2.
\end{equation}
Thus, a small phase error has only a small impact.

However, for the full network nonlocality parameters $\mathcal{F}_1$ and $\mathcal{F}_2$, for the most relevant case of $\theta=\frac{\pi}{4}$, the situation is different. Then, to second order at $\delta\approx 0$, one finds
\begin{align}
    &\mathcal{F}_1\approx \frac{1+\sqrt{2}}{2}-\frac{\delta}{2}-\frac{1+\sqrt{2}}{4}\delta^2,\qquad 
     \mathcal{F}_2\approx \frac{1+\sqrt{2}}{2}+\frac{\delta}{2}-\frac{1+\sqrt{2}}{4}\delta^2.
\end{align}
Thus, the offset is essentially  $\mathcal{F}_2-\mathcal{F}_1\approx\delta$. As we have measured $\mathcal{F}_{1}=1.112\pm0.006$ and $\mathcal{F}_{2}=1.208\pm0.006$, we have an offset of about $0.096$, which would correspond in our model here to $\delta\approx 5.5$ degrees.

\subsection{(C) Correction term to the bilocal Bell inequality~\eqref{biloc}}

\begin{figure*}[t]
\centering
\subfigure{
\includegraphics[width=0.45\textwidth]{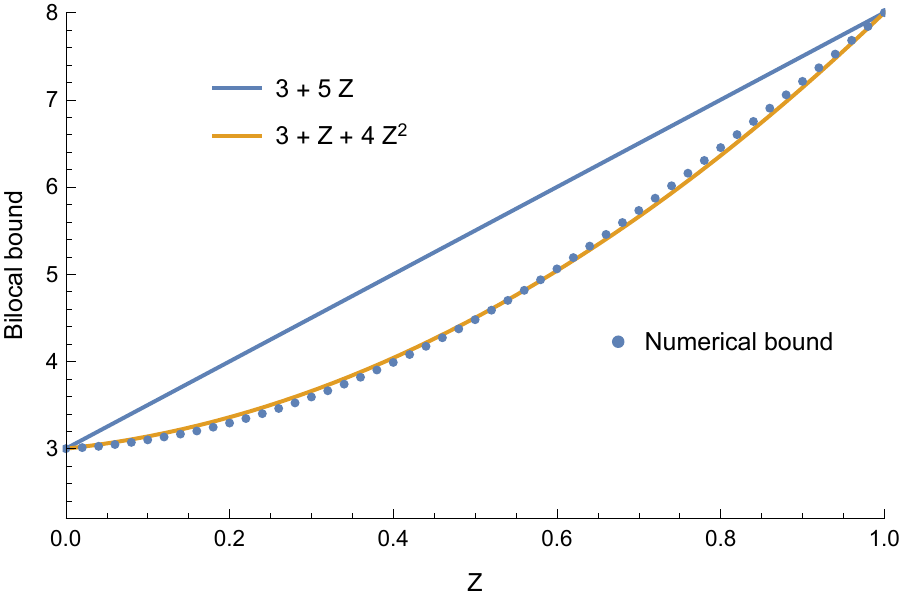}
} $\qquad$
\subfigure{
\includegraphics[width=0.45\textwidth]{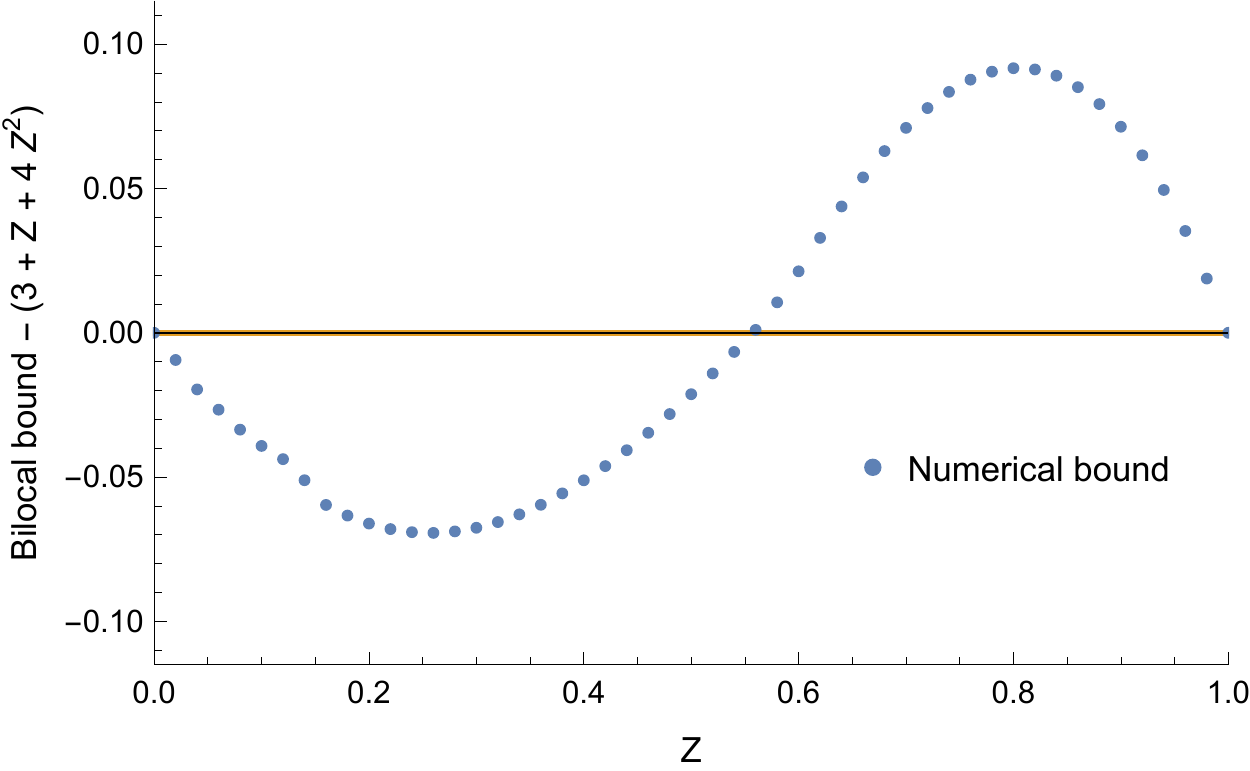}
}
\caption{Numerically obtained bilocal bounds on ${\cal B}$ for Inequality~\eqref{biloc}, versus their approximations.}
\label{fig_numerical_bnds}
\end{figure*}

The bilocal bound of Inequality~\eqref{biloc} can (reliably enough) be obtained numerically, using for instance the Fourier transform technique described in Ref.~\cite{Branciard2012}. More specifically, to obtain a bound as a function of $Z$, we imposed the linear constraints $-Z \le \langle A_1\rangle,\ldots,\langle A_3B^3C_3\rangle \le Z$ in the numerical optimization, for all correlators that do not appear in $S$ and $T$. Doing this for various values of $Z$, we obtained the values shown as blue points on Fig.~\ref{fig_numerical_bnds}~(Left).

It was already observed in Ref.~\cite{Tavakoli2021}, from these results, that taking $f(Z) = 5Z$ provides a valid, although clearly nonoptimal (for $0<Z<1$) upper bound for the inequality. For a practical test of the bilocal inequality, from experimental data that unavoidably give a nonzero value of $Z$, it is however of interest to find a tighter bound, so that it is easier to violate.

The exact bilocal bound does not seem to have a simple analytical expression. We therefore looked for simple enough functions $f(Z)$ that better approximate our numerical results and still give a valid bilocal bound.
We thus found that $f(Z)=Z+4Z^2$ provides such a tighter bound, valid (i.e., such that bilocal models remain below; see Fig.~\ref{fig_numerical_bnds}, right) as long as $Z \lesssim 0.55$---which is indeed the case in our experiment.

Note that $f(Z)=Z+4Z^2$ does still not quite match the exact bilocal bound. We use this bound for its simple expression, but note that the ``true'' violations of bilocality are in fact larger than we report in Table~\ref{tab:biloc_ineq} below.

\subsection{(D) Details of our experimental results}

\par In our experiment, we demonstrate the violation of the bilocality inequality and full network inequality  using the entanglement swapping procedure. As described in the main text, two pairs of maximally entangled states are distributed to three parties Alice, Bob and Charlie. Bob performs the Elegant Joint Measurement so that Alice and Charlie get entangled. In order to evaluate the effect of the entanglement swapping procedure, we do the state tomography of the resulting state $\rho_{AC|b}$ of Alice and Charlie, conditioned on Bob's output, and then compare it with the ideal swapped state. The average fidelities (for the different outputs of Bob) of the swapped state are as follows.

\begin{table}[thp!]
\centering
\setlength{\tabcolsep}{2mm}{
\begin{tabular}{cccccccc}
\hline\hline
$\theta$    & 0     & $\pi/12$ & $\pi/6$  & $\pi/4$  & $\pi/3$  & $5\pi/12$ & $\pi/2$  \\
Fidelity & $0.985\pm0.002$ & $0.980\pm 0.002$ & $0.975\pm0.002$ & $0.975\pm0.002$ & $0.981\pm0.001$ & $0.983\pm0.001$  & $0.988\pm0.001$ \\ \hline\hline
\end{tabular}}
\end{table}

\par We also test three different inequalities to demonstrate the EJM-based correlation. The first and second inequalities are described in detail in the main text (Eqs.~\eqref{biloc}--\eqref{fnn2}). The third inequality is given in Eq.~(11) of Ref.~\cite{Tavakoli2021}, and is also a bilocal Bell inequality tailored for the EJM. The form of this inequality is as follows
\begin{equation} \label{biloc2}
\begin{array}{c}
\mathcal{B}^{\prime} \equiv \sum_{x, b} \sqrt{p(b)\left(1-b^{x} E_{b}^{\mathrm{A}}(x)\right)}+\sum_{z, b} \sqrt{p(b)\left(1+b^{z} E_{b}^{\mathrm{C}}(z)\right)}+\sum_{x \neq z, b} \sqrt{p(b)\left(1-b^{x} b^{z} E_{b}^{\mathrm{AC}}(x, z)\right)} \leq 12\sqrt{3} + 2\sqrt{15},
\end{array}
\end{equation}
where $E_{b}^{\mathrm{A}}(x)$, $E_{b}^{\mathrm{C}}(z)$ and $E_{b}^{\mathrm{AC}}(x, z)$ are one- and two-party expectation values for Alice and Charlie (for their inputs $x,z$), conditioned on Bob’s output $b$: e.g., $E_{b}^{\mathrm{A}}(x) = \sum_{a,c} a \, P(a,c|b,x,z)$.
\par As the $\theta$ parameter changes, so does the violation of the inequalities. In the main text, we describe two special points $\theta=0$ and $\theta=\frac{\pi}{4}$ in details, which provide the maximal violations of  Eqs.~\eqref{biloc}, \eqref{biloc2}, and Eqs.~\eqref{fnn1}--\eqref{fnn2} respectively. The values of the three inequalities are listed in the following three tables.

\begin{table}[thp!]
\centering
 \begin{threeparttable}
\setlength{\tabcolsep}{1.4mm}{
\caption{\textbf{Values of the Bilocal Inequality~\eqref{biloc}}}\label{tab:biloc_ineq}
\begin{tabular}{cccccccc}
\hline\hline
$\theta$    & 0     & $\pi/12$ & $\pi/6$  & $\pi/4$  & $\pi/3$  & $5\pi/12$ & $\pi/2$ \\\hline
$\mathcal{B}$    & $3.922\pm0.018$ & $3.854\pm0.017$ & $3.700\pm0.017$ & $3.495\pm0.016$ & $3.280\pm0.015$ & $3.001\pm0.014$  & $2.861\pm0.014$ \\ 
$Z$ term              & $0.071\pm0.008$ & $0.075\pm0.008$ & $0.077\pm0.008$ & $0.074\pm0.008$ & $0.092\pm0.008$ & $0.099\pm0.008$  & $0.091\pm0.008$ \\ 
Bilocal bound       & $3.092\pm0.012$ & $3.097\pm0.012$ & $3.101\pm0.012$ & $3.093\pm0.012$ & $3.126\pm0.013$ & $3.137\pm0.013$  & $3.124\pm0.013$ \\ 
Violation           & $0.831\pm0.021$ & $0.756\pm0.022$ & $0.599\pm0.021$ & $0.399\pm0.020$ & $0.154\pm0.019$ & None   & None \\ 
NSD\tnote{1} & 39.5 & 34.4 & 28.5 & 20.1 & 8.1 &  None & None\\ \hline\hline
\end{tabular}}
\end{threeparttable}
\end{table}

\begin{table}[thp!]
\centering
\setlength{\tabcolsep}{1.4mm}{
\caption{\textbf{Values of the Full Network Inequalities~\eqref{fnn1}--\eqref{fnn2}}}
 \begin{threeparttable}
\begin{tabular}{cccccccc}
\hline\hline
$\theta$    & 0     & $\pi/12$ & $\pi/6$  & $\pi/4$  & $\pi/3$  & $5\pi/12$ & $\pi/2$ \\ \hline
$\mathcal{F}_1$    & $0.970\pm0.008$ & $1.068\pm0.007$ & $1.119\pm0.006$ & $1.112\pm0.006$ & $1.083 \pm 0.005$ & $0.982 \pm 0.005$ & $0.925 \pm 0.005 $\\
$\mathcal{F}_2$      & $1.044\pm0.008$ & $1.145\pm0.007$ & $1.200\pm0.006$ & $1.208\pm0.006$ & $1.163 \pm 0.005$ & $1.093 \pm 0.005$ & $1.011 \pm 0.005$ \\
FNN bounds  & 1 & 1 & 1 & 1 & 1 & 1 & 1 \\
Violation  & None & $0.068\pm0.007$ & $0.119\pm0.006$ & $0.112\pm0.006$ & $0.083\pm0.005$ & None & None \\ 
NSD\tnote{1} & None & 9.7 & 19.8 & 18.7 & 16.6 &  None & None\\ \hline\hline
\end{tabular}
   \end{threeparttable}}
\end{table}

\begin{table}[thp!]
\setlength{\tabcolsep}{1mm}{
\caption{\textbf{Values of the Bilocal Inequality~\eqref{biloc2}}}
\centering
\begin{threeparttable}
\begin{tabular}{cccccccc}
\hline\hline
$\theta$    & 0     & $\pi/12$ & $\pi/6$  & $\pi/4$  & $\pi/3$  & $5\pi/12$ & $\pi/2$  \\ \hline
$\mathcal{B}^{\prime}$  & $29.333\pm0.019$ & $29.163\pm0.018$ & $28.735\pm0.019$ & $28.148\pm0.019$ & $27.492\pm0.019$ & $26.750\pm0.019$ & $26.261\pm0.019$ \\
Bound     & 28.531 & 28.531 & 28.531 & 28.531 & 28.531 & 28.531 & 28.531 \\
Violation & $0.803\pm0.019$  & $0.633\pm0.018$  & $0.204\pm0.019$  & None  & None  & None  & None \\
NSD\tnote{1} & 42.3 & 35.2 & 10.7 & None & None &  None & None\\
\hline\hline
\end{tabular}
 \begin{tablenotes}
       \footnotesize
       \item[1] Number of Standard Deviations the violations correspond to.
\end{tablenotes}
   \end{threeparttable}}
\end{table}

%\newpage
\subsection{(E) Details of our EJM setup}

\begin{figure}[tph]
-\includegraphics[width=0.9\textwidth]{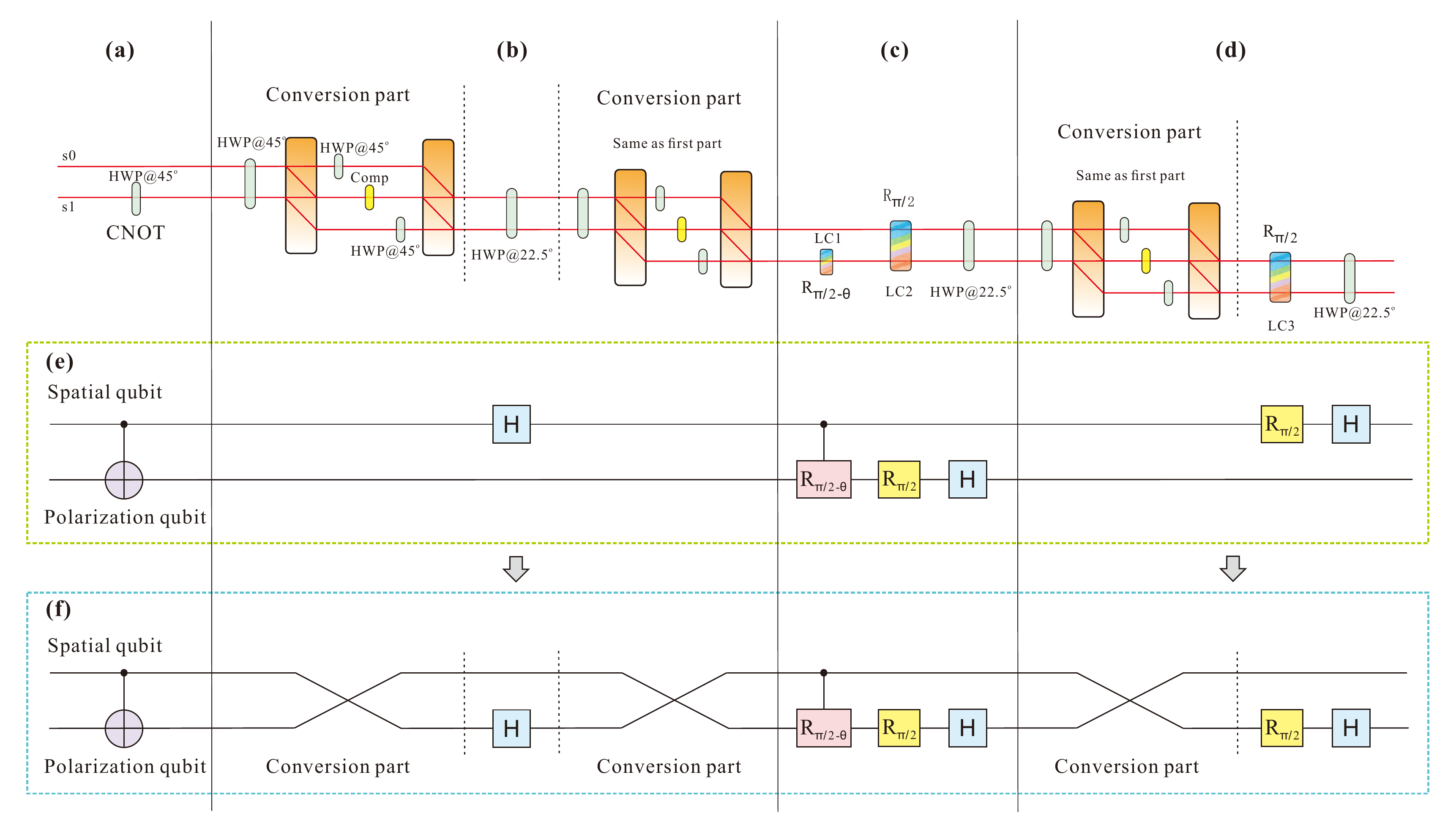}
\vspace{-0.5cm}
\caption{EJM setup. (a) CNOT gate. A HWP set at $45^{\circ}$ and placed on path $s_1$ is used to realize the CNOT gate. (In our actual experimental setup, as shown in Fig.~\ref{fig2}(d) of the main text, we combine the $45^{\circ}$ HWP of the CNOT gate with the next $45^{\circ}$ HWP in the subsequent conversion part. The $45^{\circ}$ HWP of the CNOT gate is then replaced by a $0^{\circ}$ HWP on path $s_1$; the phase it introduces is eliminated by replacing the compensator with a $0^\circ$ HWP in the following conversion part.)
(b) Hadamard operation on the spatial qubit. Two conversion parts are used to swap the spatial and polarization qubits. The Hadamard operation is then realized by a $22.5^{\circ}$ HWP operating on the polarization (that effectively encodes, at this point in the circuit, the spatial qubit). (c) Operations on the polarization qubit. The C-phase, $\frac{\pi}{2}$ phase and Hadamard operations are realized on the polarization qubit. (In the actual setup, as shown in the main text, we combine the two LCs in path $s_1$ together.) (d) Operations on the spatial qubit. We convert again the spatial qubit to polarization qubit and then use the LC3 and $22.5^{\circ}$ HWP to realize the $\frac{\pi}{2}$ phase gate and Hadamard operation on the (originally) spatial qubit. (e) Circuit diagram for the EJM (same as the main text). (f) Circuit diagram for the EJM (including the conversion part). As detailed above, we use conversion parts to swap the spatial and polarization qubits, so that all the relevant operations can be achieved on the polarization degree of freedom.}
\label{fig6}
\end{figure}

As recalled in the main text, EJMs cannot be realized in linear optics without auxiliary particles when the two qubits are encoded onto different photons~\cite{Yoran, Lutkenhaus1999,Loock2004,Tavakoli2021}.
Here we use two different degrees of freedom of the same photon to circumvent this issue. 
The central idea of our approach is to use different polarization elements (HWPs and LCs) on the two different paths of the photon, so that the spatial qubit controls the polarisation qubit. We then use polarization-to-spatial conversion setups---composed of BDs, HWPs and a compensator (used to compensate for the optical path between different paths)---to swap the two qubits ($|s_0,H\rangle\leftrightarrow|s_0,H\rangle, |s_0,V\rangle\leftrightarrow|s_1,H\rangle, |s_1,V\rangle\leftrightarrow|s_1,V\rangle$) whenever needed. See Fig.~\ref{fig6} for details.

\subsection{(F) State fidelities and source independence}

In our experiment, we do a 4-qubit tomography \cite{Hu:20} to estimate the fidelity and quantify the independence of the hyper-entanglement source. Using the maximum likelihood estimation, we can reconstruct the density matrix $\rho$ of the 4-qubit entangled state, and then compare it with the ideal target state $|\phi^+\rangle\otimes|\phi^+\rangle$. As already given in Appendix (A), the fidelity of the 4-qubit state is found to be $99.0\pm0.01\%$. We also calculate the fidelity of the spatial and polarization qubit pairs respectively. By tracing the path (polarization) degree of freedom, we can obtain the reduced density matrix of the polarization (spatial) qubit pair.  The fidelities of the reduced states are $0.992\pm0.01\%$ (polarization qubits $\rho_{AB_1}$) and $0.993\pm0.01\%$ (spatial qubits $\rho_{B_2C}$), respectively. 

In order to estimate the independence between the two degrees of freedom, we have evaluated the fidelity between the tomographic reconstruction of $\rho$ and the product of its reductions to systems $AB_1$ and $B_2C$, as well as their quantum mutual information~\cite{RevModPhys.74.197}, i.e., the quantum relative entropy $S(\rho||\rho_{AB_1}\otimes\rho_{B_2C})=\textup{Tr}[\rho( \log_2 \rho - \log_2(\rho_{AB_1}\otimes\rho_{B_2C}))]$.
The results are $99.1\pm 0.1\%$ for the fidelity and $0.048\pm 0.003$ for the mutual information, which indeed shows that our input state is  close to a product state and that the polarization and spatial qubit pairs are close to being independent.

\end{document}